\documentclass[aps,prl,twocolumn,superscriptaddress,showpacs,preprintnumbers,amsmath,amssymb,nofootinbib]{revtex4-1}
\usepackage[none]{hyphenat}

\usepackage{graphicx} 
\usepackage{dcolumn}  

\graphicspath{{ps}}

\newcommand{\ee}{e^{+}e^{-}}

\newcommand{\jp}{J/\psi}

\newcommand{\psip}{\psi^{\prime}}

\newcommand{\pipi}{\pi^{+}\pi^{-}}

\newcommand{\ccbar}{c\bar{c}}

\newcommand{\rt}{\rightarrow}

\newcommand{\etal}{\em et al.}
\newcommand{\jpsi}{J/\psi}

\newcommand{\chicp}{\chi_{c0}^{\prime}}
\newcommand{\chictwop}{\chi_{c0}(2P)}

\begin{document}


\preprint{\vbox{ }} 

\title{ \quad\\[0.5cm] Is the ${\boldmath X(3915)}$ the ${\boldmath \chictwop}$? }

\author{Stephen Lars Olsen}\affiliation{Center for Underground Physics, Institute for Basic Science, Daejeon 305-811, Korea}

\date{\today}


\begin{abstract}
The Particle Data Group has assigned the $X(3915)$ meson, an $\omega\jp$ mass peak seen
in $B\rt K \omega\jp$ decays and $\gamma\gamma \rt\omega\jp$ two-photon fusion
reactions, as the $\chi_{c0}(2P)$, the $2^3P_0$ charmonium state.  Here it is shown that
if the $X(3915)$ is the $\chi_{c0}(2P)$, the measured strength of the
$\gamma\gamma\rt X(3915)$ signal implies an {\it upper} limit on the branching fraction
${\mathcal B}(\chi_{c0}(2P)\rt\omega\jp)<7.8\%$ that conflicts with a $>14.3\%$ {\it lower}
limit derived for the same quantity from the $B\rt K X(3915)$ decay rate.
Also,  the absence any signal for $X(3915)\rt D^0\bar{D}^0$ in $B^+\rt K^+ D^0\bar{D}^0$
decays is used to establish the limit
${\boldmath B}(X(3915)\rt D^0\bar{D}^0 < 1.2\times {\mathcal B}(X(3915)\rt \omega\jp)$.  This
contradicts expectations that $\chi_{c0}(2P)$ decays to $D^0\bar{D}^0$ would
be a dominant process, while decays to $\omega\jp$, which are Okubo-Zweig-Iizuka
suppressed, would be relatively rare.  These, plus reasons given earlier by Guo and
Meissner, raise serious doubts about the $X(3915)=\chi_{c0}(2P)$ assignment.
\end{abstract}


\maketitle


{\renewcommand{\thefootnote}{\fnsymbol{footnote}}}
\setcounter{footnote}{0}

\section{Introduction}
\noindent

A number of meson candidates, dubbed the $XYZ$ mesons, that contain charmed-quark
anticharmed-quark ($\ccbar$)
pairs but do not match expectations for any of the unassigned levels of the $\ccbar$ charmonium
spectrum have been observed in recent experiments.  Some have non-zero electric
charge~\cite{z_c} and cannot be accommodated in the spectrum of charmonium mesons, which
are all electrically neutral.  Others are neutral and have quantum numbers that are
accessible by $\ccbar$ systems, but have properties that fail to match the tightly constrained
expectations of any of the unassigned charmonium states~\cite{xyz}.  To date, there is no compelling
theoretical explanation for these $XYZ$ mesons.  Experimental observations of additional
states and more refined measurements of properties of the existing states may eventually
reveal patterns that give clues to their underlying structure.  An important part of this
program is a careful distinction of new states that are conventional charmonium mesons from those
that are not.

The $X(3915)$ was observed by Belle as a near-threshold peak in the $\omega\jpsi$
invariant mass distribution in exclusive $B\rt K\omega\jp$ decays~\cite{belle_y3940};
it was subsequently confirmed  by BaBar~\cite{babar_y3940}.  An $\omega\jp$ mass peak with
similar mass and width was reported by Belle in the two-photon fusion process
$\gamma\gamma\rt\omega\jp$ in 2010~\cite{belle_x3915}.   BaBar reported confirmation of the 
$\gamma\gamma\rt X(3915)\rt\omega\jp$ observation~\cite{babar_x3915} and, from a
study of the angular correlations among the final-state particles, established
the $J^{PC}$ quantum numbers to be $0^{++}$. 

The similar masses and widths of the peaks seen in $B$ decay and in two-photon
fusion processes suggest that these are two different production mechanisms for the same
state. The Particle Data Group's (PDG) average values of the mass and width are~\cite{pdg}:
\begin{eqnarray}
     M(X(3915)) &=& 3918.4\pm 1.9~{\rm MeV} \nonumber \\
\Gamma(X(3915)) &=& 20.0 \pm 5.0~{\rm MeV}.
\end{eqnarray}
The weighted average of the Belle~\cite{belle_y3940} and BaBar~\cite{babar_y3940} product 
branching fraction measurements for $X(3915)$ production in $B$ decay is 
\begin{eqnarray} 
{\mathcal B}(B^+\rt K^+ X(3915))\times {\mathcal B}(X(3915)\rt\omega\jp)\nonumber\\
= 3.2\pm 0.9\times 10^{-5},  
\label{eqn:b2kx}
\end{eqnarray}
while the average of measured production rates in two-photon fusion (using $J^{PC}=0^{++}$)
gives~\cite{pdg}
\begin{equation}
\Gamma_{X(3915)}^{\gamma\gamma}\times {\mathcal B}(X(3915)\rt \omega\jp) = 54\pm 9 {\rm eV},
\label{eqn:c0-ggwidth}
\end{equation}
where $\Gamma_{X(3915)}^{\gamma\gamma}$ is the partial width for $X(3915)\rt\gamma\gamma$.

The presence of a $\jp$ among its decay products indicate that the $X(3915)$ contains
a $\ccbar$ quark pair.  The only unassigned $0^{++}$ charmonium level in the vicinity
of the $X(3915)$ mass is the $\chictwop$, the first radial exitation of the $\chi_{c0}$
charmonium state.  (In the following, the $\chictwop$ referred to as the $\chicp$.) 
Because of this, the PDG identifies the $X(3915)$ as the $\chicp$.
This assignment was disputed by Guo and Meissner~\cite{guo}, primarily because:
\begin{itemize}
\item
the partial width for $X(3915)\rt \omega\jp$ is too large for a decay process that
is Okubo-Zweig-Iizuka (OZI) suppressed for a charmonium state;
\item
the lack of evidence for $X(3915)\rt D\bar{D}$ decays, which are expected to be dominant
$\chicp$ decay modes;
\item
the small $\chi_{c2}(2P)$-$\chictwop$ mass splitting.  
\end{itemize}

If the $X(3915)$ is not conventional charmonium but, instead, another $XYZ$ meson,
it would be the lightest observed scalar and one of the narrowest of the new states.
As such, it would likely play a key role in in attempts to understand their underlying nature.
Thus, the validity of the PDG assignment
of  the $X(3915)$ as the $\chicp$ is a critical issue that needs to be carefully addressed.
In this report I amplify some of the Guo-Meissner concerns and identify some other serious problems
with the $X(3915)=\chicp$ assignment.

\section{The {\boldmath $\chi_{c2}(2P)$} charmonium state}
\begin{figure}[htb]
  \includegraphics[height=0.3\textwidth,width=0.45\textwidth]{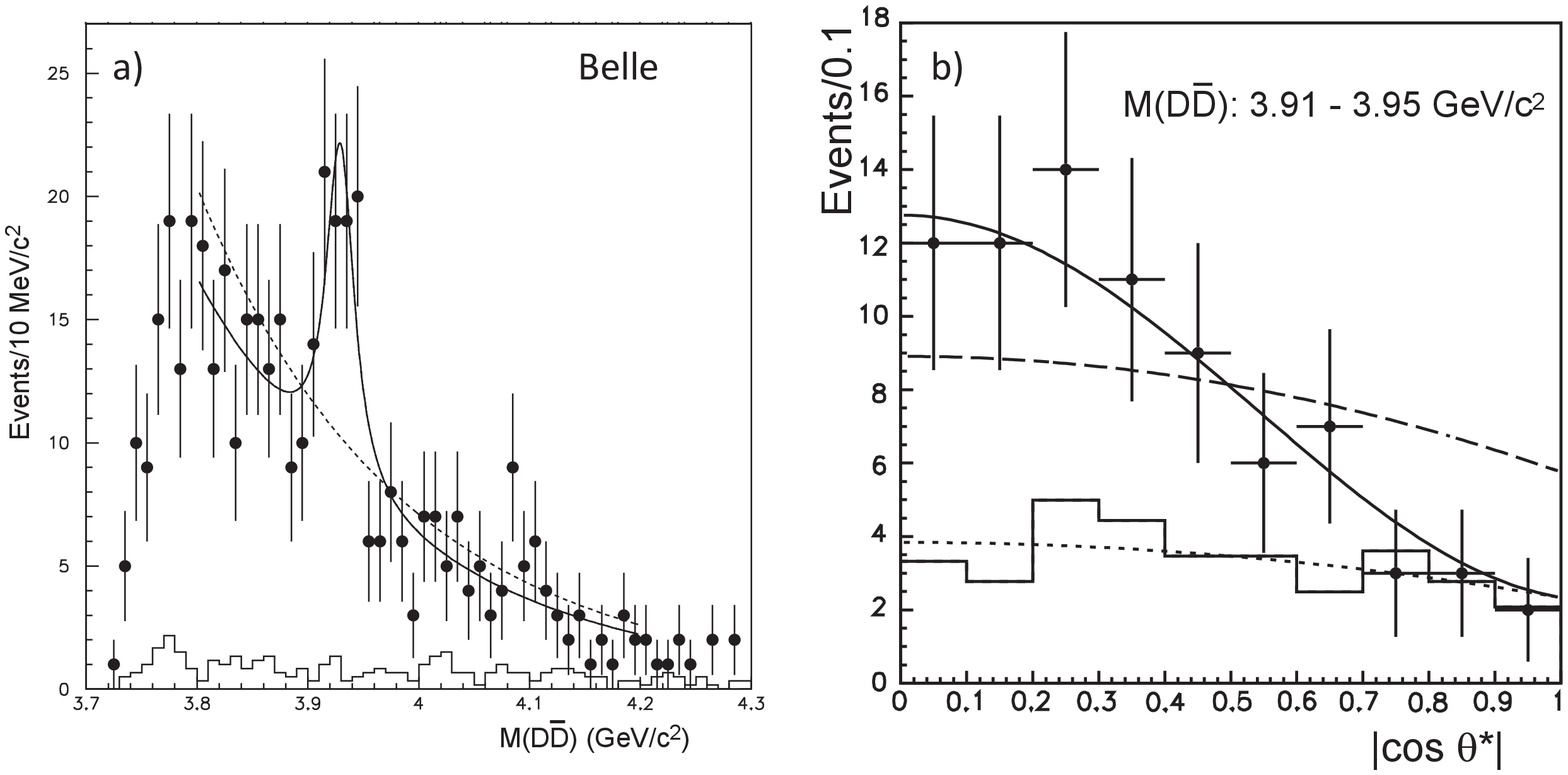}
\caption{\footnotesize {\bf a)} The $M(D \bar{D})$ distributions for 
$\gamma\gamma\rt D\bar{D}$ decays from ref.~\cite{belle_z3930}.  The open histogram
shows the background level determined from $D$ mass sidebands.  The solid (dashed) curve
shows the result of a fit that includes (excludes) a $\chi_{c2}^{\prime}$ signal.
{\bf b)} The $dN/d|\cos\theta^*|$ distribution for events in the peak region.  The
solid (dashed) curve shows expectations for $J=2$  ($J=0$).
The histogram shows the non-resonant contribution.
}
\label{fig:z3930}
\end{figure}

The properties of the $\chicp$ are constrained by measurements of its $J=2$ multiplet partner,
the $\chi_{c2}(2P)$, or $\chi_{c2}^{\prime}$, that was
seen by both Belle~\cite{belle_z3930} and BaBar~\cite{babar_z3930} as a distinct
$M(D\bar{D})$ peak in the two-photon fusion process $\gamma\gamma\rt D\bar{D}$. Both
groups see a clear $\sin^4\theta^*$ production angle dependence that is characteristic
of a $J=2$ charmonium state, and there are no reasons to question the $\chi_{c2}^{\prime}$
assignment. The Belle $M(D\bar{D})$ and $dN/d|\cos\theta^*|$ distributions are shown in
Fig.~\ref{fig:z3930}.  Belle and BaBar measurements for the the mass and width are in good
agreement; the PDG average values are~\cite{pdg}:
\begin{eqnarray}
     M(\chi_{c2}^{\prime}) &=& 3927.2\pm 2.6~{\rm MeV} \nonumber \\
\Gamma(\chi_{c2}^{\prime}) &=& 24.0 \pm 6.0~{\rm MeV}.
\end{eqnarray}
Belle and BaBar measurements of its two-photon production rate are also in good agreement
and are characterized by the product~\cite{pdg}
\begin{equation}
\Gamma_{\chi_{c2}^{\prime}}^{\gamma\gamma}\times {\mathcal B}(\chi_{c2}^{\prime}\rt D\bar{D}) = 210\pm 40 {\rm eV}.
\label{eqn:c2-ggwidth}
\end{equation}

\section{Consequences of {\boldmath $X(3915)=\chicp$}}

\subsection{The {\boldmath $\chi_{c2}(2P)-\chictwop$} mass splitting}
As pointed out by Guo and Meissner, the $X(3915)=\chicp$ assignment implies
an anomalously small $\chi_{c2}(2P)-\chictwop$ mass splitting.  Current
measurements put it at $\Delta M(2P)=8.8\pm 3.2$~MeV, in which case 
\begin{equation}
r_c\equiv \frac{\Delta M(2P)}{\Delta M(1P)}= 0.06\pm 0.02.
\label{eqn:split}
\end{equation}
This is much smaller than the corresponding ratio for the bottomonium
system, $r_b=0.69\pm 0.01$, and potential model predictions for charmonium
that are in the range $ 0.6 < r_c <0.9$~\cite{barnes}.  A study of 
modifications to potential-model mass calculations caused by couplings to
open charmed mesons, finds that these effects reduce the expected splitting to 
$r_c\simeq 0.24$~\cite{eichten}, which is still large
compared to the observed splitting (Eq.~\ref{eqn:split}).  Moreover, this
study used a a stronger $\chicp$-$D\bar{D}$ coupling strength then can be
supported by measured $X(3915)$ data; the study's resultant $\chicp\rt D\bar{D}$
partial width was 12.4~MeV, which is nearly twice
as large as an upper limit on this quantity that is derived below.

\subsection{Limits on {\boldmath ${\mathcal B}(\chicp \rt \omega\jp)$}}

Using measured numbers and some conservative assumptions, I derive an upper limit on
${\mathcal B}(\chicp\rt \omega\jp)$ from the two-photon fusion production rate that is
more than a factor of two below a lower limit on the same quantity determined from the
rate for $\chicp$ production in $B^+\rt K^+\chicp$ decays.  

\paragraph{From $\gamma\gamma\rt\chicp \rt \omega\jp$:}~~~
From Eq.~\ref{eqn:c0-ggwidth} it is clear that an {\it upper} limit on ${\mathcal B}(\chicp\rt\omega\jp)$
can be inferred from a lower limit on $\Gamma_{\chicp}^{\gamma\gamma}$.  Potential model relations for
$\Gamma_{\chi_{c0}(nP)}^{\gamma\gamma}$ and $\Gamma_{\chi_{c2}(nP)}^{\gamma\gamma}$ with one-loop QCD
corrections are~\cite{ggwidth}:
\begin{eqnarray}
\Gamma_{\chi_{c0}(nP)}^{\gamma\gamma} &=&
\frac{27e_c^4\alpha^2}{\mu^4}|\frac{d R_n}{d r}(0)|^2
\lbrack 1+\frac{\alpha_s}{\pi} (\frac{\pi^2}{3}-\frac{28}{9} ) \rbrack \nonumber \\
\Gamma_{\chi_{c2}(nP)}^{\gamma\gamma} &=&
\frac{36e_c^4\alpha^2}{5\mu^4}|\frac{d R_n}{d r}(0)|^2
\lbrack 1-\frac{\alpha_s}{\pi}\frac{16}{3} \rbrack,
\label{eqn:digamma}
\end{eqnarray}
where $e_c=2/3$ is the charmed quark charge, $\alpha$ is the fine structure constant,
$\mu$ is the reduced charmed quark mass, $R_n(r)$ is the $\chi_{cJ}(nP)$ radial wave function
and $\alpha_s$ is the QCD coupling
strength.  From Eq.~\ref{eqn:digamma} one can infer the relation
\begin{equation}
\frac{\Gamma_{\chicp}^{\gamma\gamma}}{\Gamma_{\chi_{c2}^{\prime}}^{\gamma\gamma}} \simeq
\frac{\Gamma_{\chi_{c0}}^{\gamma\gamma}}{\Gamma_{\chi_{c2}}^{\gamma\gamma}} =4.5\pm 0.6,
\label{eqn:c2c0}
\end{equation}
which is valid in potential models to the level of changes in the QCD correction factors
due to the running of $\alpha_s$ between the $\chi_{cJ}$ and $\chi_{cJ}^{\prime}$ masses,
which are a few percent.

Since  ${\mathcal B}(\chi_{c2}^{\prime}\rt D\bar{D})$ is necessarily less than unity, Eq.~\ref{eqn:c2-ggwidth}
implies a lower bound on $\Gamma_{\chi_{c2}^\prime}^{\gamma\gamma}$.  
This, together with Eqs.~\ref{eqn:c2c0} and \ref{eqn:c0-ggwidth}, translates
into the (90\% CL) upper limit\footnote{In this paper I assume all errors are Gaussian
and, when required, combine statistical and systematic errors in quadrature.}
\begin{equation}
{\mathcal B}(\chicp \rt \omega\jp)<7.8\%.
\label{eqn:wjp-upper}
\end{equation}

\paragraph{From $B^+\rt K^+ \chicp$; $\chicp \rt \omega\jp$:}~~~
A lower limit of ${\mathcal B}(\chicp\rt\omega\jp)$ can be deduced from Eq.~\ref{eqn:b2kx}
if an upper limit on ${\mathcal B}(B^+\rt K^+\chicp)$ can be established.  Here I assume
that the ${\mathcal B}(B^+\rt K^+\chicp)$ is less than or equal to
${\mathcal B}(B^+\rt K^+\chi_{c0})$, where the latter branching fraction has been
measured to be~\cite{pdg} 
\begin{equation}
{\mathcal B}(B^+\rt K^+\chi_{c0}) = 1.34^{+0.19}_{-0.16}\times 10^{-4}.
\label{eqn:b2kchicp}
\end{equation}
This assumption is reasonable for a few reasons, including: the available phase space for
$B\rt K\chicp$  is significantly smaller than that for $B\rt K\chi_{c0}$; the $B$-meson
decay rate to $P$-wave charmonium mesons is expected to be 
proportional to the $|\frac{d R_n}{d r}(0)|^2$~\cite{bodwin}, which
decreases with increasing $n$.  Moreover, measured $B$ meson branching fractions to
excited charmonium states, where they exist, are all smaller than those to the ground
states.\footnote{
For example~\cite{pdg}, 
${\mathcal B}(B^+\rt K^+\psip)/{\mathcal B}(B^+\rt K^+\jp) = 0.63\pm 0.04$,   
${\mathcal B}(B^+\rt K^+\eta_c(2S))/{\mathcal B}(B^+\rt K^+\eta_c) = 0.35\pm 0.19$ and
${\mathcal B}(B^+\rt K^*(890)^+\psip)/{\mathcal B}(B^+\rt K^*(890)^+\jp) = 0.47\pm 0.10$.}
With this assumption, Eq.~\ref{eqn:b2kx} translates into a 90\% CL {\it lower} limit of
\begin{equation}
{\mathcal B}(\chicp \rt \omega\jp)>14.3\%,
\label{eqn:wjp-lower}
\end{equation}
which is nearly a factor of two higher than the {\it upper} limit given in Eq.~\ref{eqn:wjp-upper}.

\subsection{Limits on {\boldmath ${\mathcal B}(\chicp \rt D\bar{D})$}}

A peculiar feature of the $X(3915)$ is the absence of any evidence for it in 
either $\gamma\gamma \rt D\bar{D}$  or $B\rt K D\bar{D}$ processes even though
Belle and BaBar have each studied both channels.  Here I discuss limits on
${\mathcal B}(\chicp \rt D\bar{D})$ from these processes.

\paragraph{ From $\gamma\gamma\rt X(3915) \rt D\bar{D}$:}~~~
The possibility that an $X(3915)\rt D\bar{D}$ signal is lurking in the
$M(D\bar{D})$ and $dN/d|\cos\theta^*|$ distributions  for $\gamma\gamma\rt D\bar{D}$ from
Belle and BaBar was examined by Chen, He, Liu and Matsuki~\cite{matsuki}.  Based on fits
to Belle's measurements, they claim a signal for  $\gamma\gamma\rt X(3915)\rt D\bar{D}$
with marginal significance  at
a strength that is 69\% of that for the $\chi_{c2}^{\prime}$.  I use this result to conclude
that  $\Gamma_{X(3872)}^{\gamma\gamma}\times {\mathcal B}(X(3915) \rt D\bar{D}) <
\Gamma_{\chi_{c2}^{\prime}}^{\gamma\gamma}\times {\mathcal B}(\chi_{c2}^{\prime}\rt D\bar{D})$.
If $X(3915)=\chicp$, this, together with the ratio given
in Eq.~\ref{eqn:c2c0}, implies a 90\% CL upper limit
\begin{equation}
{\mathcal B}(\chicp\rt D\bar{D})<0.25{\mathcal B}(\chi_{c2}^{\prime}\rt D\bar{D}) < 25\%,
\end{equation}
which is well below theoretical expectations of nearly 100\%~\cite{barnes,eichten}.
This upper limit corresponds to an upper limit on the partial width for
$\chicp\rt D\bar{D}$ of 7~MeV.

\begin{figure}[htb]
  \includegraphics[height=0.35\textwidth,width=0.35\textwidth]{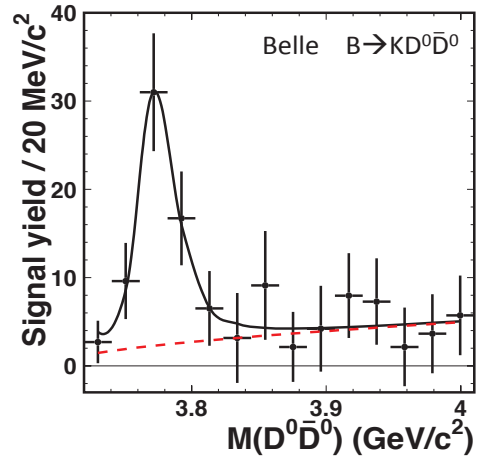}
\caption{\footnotesize The $M(D^0 \bar{D}^0)$ distribution for
$B\rt K D^0\bar{D}^0$ decays from ref.~\cite{belle_kd0d0}.  The peak near $3.77$~GeV
is due to the $\psi(3770)$. 
}
\label{fig:d0d0}
\end{figure}

\paragraph{From $B^+\rt K^+  X(3915)$, $X(3915) \rt D^0\bar{D}^0$:}~~~
The $D^0\bar{D}^0$ invariant mass distribution for $B^+\rt K^+ D^0\bar{D}^0$ decays
from the Belle experiment~\cite{belle_kd0d0} is shown in Fig.~\ref{fig:d0d0}, where
a strong, $68\pm 15 $ event signal for $B^+\rt K^+\psi(3770)$; $\psi(3770)\rt D^0\bar{D}^0$
is evident.  The measured product branching fraction for this process is~\cite{pdg}:
\begin{eqnarray}
{\mathcal B}(B^+\rt K^+\psi(3770))\times {\mathcal B}(\psi(3770)\rt D^0\bar{D}^0)\nonumber \\
= 1.6\pm 0.6 \times 10^{-4}.
\end{eqnarray}
There is no sign in Fig.~\ref{fig:d0d0} of a peak near $M(D^0\bar{D}^0)\sim 3.92$~GeV
that would correspond to the decay chain $B^+\rt K^+ X(3915)$, $X(3915)\rt D^0\bar{D}^0$.
In fact, the Belle analysis attributes most of the events that are seen in the 3.92~GeV mass
region to the process $B^+\rt D_{sJ}(2700)^+ \bar{D}^0$; $D_{sJ}(2700)^+\rt K^+ D^0$. Ignoring
this possibility and attributing all of the $8\pm 5$ events in the 20~MeV-wide bin centered
at 3.917~GeV to the $X(3915)$, then scaling this to the $\psi(3770)$ signal (assuming
constant acceptance) and comparing the result with the measured rate for $X(3915)\rt\omega\jp$
production in $B$ decays (Eq.~\ref{eqn:b2kx}) gives the 90\% CL limit
\begin{equation}
{\mathcal B}(X(3915)\rt D^0\bar{D}^0)< 1.2\times {\mathcal B}(X(3915)\rt \omega\jp),
\end{equation}
which is independent of the $X(3915)=\chicp$ assumption or any properties of the charmonium
model. For the $X(3915)=\chicp$ scenario, this strongly conflicts with theoretical expectations
that $\chicp\rt D^0\bar{D}^0$ should be a dominant ``fall-apart'' mode with a branching fraction
that is nearly 50\%~\cite{eichten}, while $\chicp\rt\omega\jp$ would be an OZI-suppressed
decay mode.  Measured OZI-suppressed charmonium decays have partial widths that are of order
100~keV or less.  A partial width of this magnitude would correspond to a $\chicp\rt\omega\jp$
branching fraction that is below 1\%.

\section{Discussion}

Any one of the points raised above would make the $X(3915)=\chicp$ assignment unlikely;
the combination of them all provides a {\it prima facie} case that it is incorrect or,
at a minimum, premature.  On the other hand, if the $X(3915)$ is not the $\chicp$, what is
it? Also, where is the real $\chicp$? In the following, I briefly discuss these issues.

\subsection{Where is the {\boldmath $\chicp$}?}

Gou and Meissner suggested that the ``non-resonant'' events seen by Belle and
BaBar in the $\gamma\gamma\rt D\bar{D}$ distribution (see Fig.~\ref{fig:z3930}a)
are, in fact, due to $\chicp\rt D\bar{D}$.  In their paper~\cite{guo}, they present
fits to both the Bellle and BaBar data that include a broad Breit Wigner (BW) function
to represent the events under the $\chi_{c2}^{\prime}$ peak.  Their fits to both
experiments' distributions give an average  mass and width for the broad BW of
$M=3837.6\pm 11.5$~MeV and $\Gamma = 221\pm 19$~MeV that they attribute to the $\chicp$. 
This mass, the strong $D\bar{D}$ signal, and the absence of any sign of a similarly broad
signal in the $\gamma\gamma\rt\omega\jp$ distributions from Belle~\cite{belle_x3915} and
BaBar~\cite{babar_x3915}, insure that this candidate for the $\chicp$ does not suffer
from any of the difficulties listed above.   However, in their fit, Guo and Meissner
ignore possible feeddown to the $D\bar{D}$ from $\chi_{c0}^{\prime}\rt D\bar{D}^*$; 
$\bar{D}^*\rt D\pi (\gamma)$, where the $\pi$ or $\gamma$ is undetected.
The $D\bar{D}$ events from this process would concentrate in a region that is about one pion
mass below the $\chi_{c0}^{\prime}$ peak and below the peak value from their fit.
Since theoretical estimates of the $\chi_{c0}^{\prime}\rt D\bar{D}^*$ give a rate 
that is in the range of $0.3\sim 0.5$ times that for $D\bar{D}$~\cite{barnes,eichten}, 
this background would likely bias the Guo-Meissner fit to a lower mass value.  With enough
events, the strength of the $D\bar{D}^*$ contribution could be determined from the
number of $D^+\bar{D}^0$ events in the data sample and a combined study of neutral
and charged $D\bar{D}^*$ pairs could remove such a bias~\cite{uehara}.  This could be done at
BelleII~\cite{belle2}.

\begin{figure}[htb]
  \includegraphics[height=0.2\textwidth,width=0.45\textwidth]{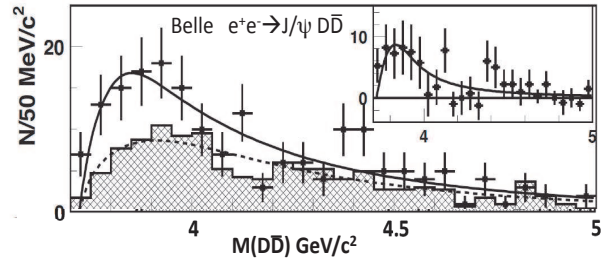}
\caption{\footnotesize The $M(D \bar{D})$ distribution for
$\ee \rt \jp D\bar{D}$ decays from ref.~\cite{belle_x4160}.  The shaded histogram
is the sum of non-$\jp$ and non-$D\bar{D}$ backgrounds.  The inset shows the
background-subtracted distribution. The solid curve is the result of the fit
discussed in the text. 
}
\label{fig:chicp}
\end{figure}

A Belle study of the annihilation processes $\ee\rt\jp (\ccbar)$, where $(\ccbar)$
represents charmonium states, found significant cross sections only for cases where
$(\ccbar)$ has zero spin~\cite{belle_x3940}.  This suggests that a signal for the
$\chicp$ might show up in $\ee\rt\jp D\bar{D}$ events.  Figure~\ref{fig:chicp} shows
the $D\bar{D}$ invariant mass distribution for this process~\cite{belle_x4160}.
Here there is a clear excess of events above the non-$\jp$ and/or non-$D\bar{D}$ backgrounds
that are reliably determined from $\jp$ and $D$ mass sideband data and shown as a hatched
histogram.  Chao suggested that this event excess is due to the $\chicp$~\cite{chao}. The
Belle fit to this excess, shown in the figure as a solid curve, returned a signal with a
statistical significance of $3.8\sigma$ and a mass and width of $M=3878\pm 48$~MeV and
$\Gamma=347 ^{+316}_{-143}$~MeV, which are consistent with the Guo-Meissner fit values
discussed above.  However, since the Belle fit was unstable under variations of the
background parameterization and and the bin width, they made no claims
for an observation.  A reanalysis of this channel with a larger data sample by the Belle group
is currently in progress~\cite{mizuk}.

\subsection{What is the {\boldmath $X(3915)$}?}

A variety of structures for the $XYZ$ mesons have been suggested, including:
quark-antiquark-gluon hybrids~\cite{hybrids};  tightly bound QCD tetraquarks in colored
diquark-diantiquark configurations~\cite{diquarks}; molecule-like structures formed from
mesons bound by nuclear-like meson-exchange forces~\cite{molecules}; and hadrocharmonium
in which $\ccbar$ states are bound to light-quarks and/or gluons via
chromo-electric dipole forces~\cite{hadrocharmonium}.
In addition, some signals attributed to $XYZ$ states have been interpreted as being due to
cusps produced by coupled-channel near-threshold dynamics involving open-charmed
mesons~\cite{bugg}. 

The $X(3915)$ mass is well below Lattice QCD calculated values for the lightest $0^{++}$
charmonium hybrid, which are around 4450~MeV~\cite{dudek}. It is also far from any
relevant open-charmed-meson threshold.  Thus, hybrid and cusp interpretations for the
$X(3915)$ can probably be ruled out.   

For a four-quark substructure, either of the QCD-tetraquark or meson-antimeson molecule
variety, the decay $X(3915)\rt\omega\jp$ would not be OZI-suppressed.  In the QCD-tetraquark
picture, the charmed and anticharmed quarks have correlated colors and are in close spatial
proximity, conditions that could facilitate decays to hidden charmonium states. In a detailed
study of one of the $XYZ$ states (the $Z_c(3900)$) in the context of a QCD-tetraquark picture,
partial widths for hadronic decays to hidden-charm states are found to be much larger than
those for decays to open-charmed mesons~\cite{maiani_z3900}.  The hadrocharmonium model predicts
similar results~\cite{voloshin_z3900}. One could expect similar results from a
corresponding analyses of the $X(3915)$.

If, instead, the $X(3915)$ has a $D^*\bar{D}^*$ molecule-like configuration, the charmed and
anticharmed quarks would have less spatial overlap and their colors would be uncorrelated. In this
case, one might expect that even though decays to final states with hidden charmonium are not
OZI-suppressed, they might still not be very prominent.  However, a specific model by Molina
and Oset finds a $0^{++}$ (mostly) $D^*\bar{D}^*$ bound state with mass and width $M=3943$~MeV and
$\Gamma =17$~MeV that couples strongly to $\omega\jp$ and weakly to
$D\bar{D}$~\cite{molina_x3915}, properties that match reasonably well to those of the $X(3915)$.
This model predicts significant decays to $\phi\phi$ and $\omega\omega$ final states.
 
The $X(3915)$ could probably be accommodated by some versions of either a QCD-tetraquark model or
a  molecule-like picture.  More experimental information about other decay channels, especially
$D\bar{D}$, $\pipi\chi_{c0}$, $\omega\omega$ and $\phi\phi$  $\pipi$, and searches for other,
possibly related states in the $3900$~MeV mass region might help distinguish between the two
possibilities.

\section{Summary}

The mass, production rates and limits on the $\omega\jp$ and $D\bar{D}$ branching fractions
of the $X(3915)$ make it a poor candidate for the $\chicp$ charmonium state.   

\section{Acknowledgments}

This work was supported by the Korean Insitute for Basic Science under project code IBS-R016-D1.


\begin{thebibliography}{99}



\bibitem{z_c}
S.-K. Choi {\etal} (Belle Collaboration),
Phys. Rev. Lett. {\bf 100}, 142001 (2008);
R.~Mizuk {\etal} (Belle Collaboration),
Phys. Rev. D {\bf 78}, 072004 (2008);
M.~Ablikim {\it et al.} (BESIII Collaboration),
Phys. Rev. Lett. {\bf 110}, 252001 (2013);
Z.Q.~Liu {\etal} (Belle Collaboration),
Phys. Rev. Lett. {\bf 110}, 252002 (2013);
M.~Ablikim {\etal} (BESIII Collaboration),
Phys. Rev. Lett. {\bf 111}, 242001 (2013);
M.~Ablikim {\etal} (BESIII Collaboration),
Phys. Rev. Lett. {\bf 112}, 022001 (2014);
M.~Ablikim {\etal} (BESIII Collaboration),
Phys. Rev. Lett. {\bf 112}, 132001 (2014).

\bibitem{xyz}
For a recent review see N.~Brambilla {\etal},
Eur. Phys. J. C {\bf 71}, 1534 (2011). 

\bibitem{belle_y3940}S.-K. Choi {\etal} (Belle Collaboration),
Phys. Rev. Lett. {\bf 94}, 182002 (2005).

\bibitem{babar_y3940}
P.~del~Amo~Sanchez {\it et al.} (BaBar Collaboration),
Phys. Rev. D {\bf 82}, 011101R (2010) and 
B.~Aubert {\it et al.} (BaBar Collaboration),
Phys. Rev. Lett. {\bf 101}, 082001 (2008).

\bibitem{belle_x3915} S.~Uehara {\etal} (Belle Collaboration),
Phys. Rev. Lett. {\bf 104}, 092001 (2010).

\bibitem{babar_x3915} J.P.~Lees {\it et al.} (BaBar Collaboration),
Phys. Rev. D {\bf 86}, 072002 (2012).

\bibitem{pdg} K.A.~Olive {\it et al.} (Particle Data Group),
 Chin. Phys. C {\bf 38}, 090001 (2014).


\bibitem{guo} F.-K.~Guo and U.-G.~Meissner
Phys. Rev. D {\bf 86}, 091501 (2012).

\bibitem{belle_z3930} S.~Uehara {\etal} (Belle Collaboration),
Phys. Rev. Lett. {\bf 96}, 082003 (2006).

\bibitem{babar_z3930}
B.~Aubert {\it et al.} (BaBar Collaboration),
Phys. Rev. D {\bf 81}, 092003 (2010).

\bibitem{barnes}
T.~Barnes, S.~Godfrey and E.S.~Swanson,
Phys. Rev. D {\bf 72}, 054026 (2005).

\bibitem{eichten}
E.J.~Eichten, K.~Lane and C.~Quigg,
Phys. Rev. D {\bf 69}, 094019 (2004).

\bibitem{ggwidth} 
H.W.~Huang, C.F.~Qiao and K.T.~Chao,
arXiv:hep-ph/0109054;
H.W.~Huang, C.F.~Qiao and K.T.~Chao,
Phys. Rev. D {\bf 54}, 2123 (1996);
A.~Petrelli, M.~Cacciari, M.~Greco, F.~Maltoni and M.L.~Mangano,
Nucl. Phys. B {\bf 514}, 245 (1998);
R.~Barbieri, M.~Caffo, R.~Gatto and E.~Remiddi,
Nucl. Phys. B {\bf 192}, 61 (1981);
R.~Barbieri, M.~Caffo, R.~Gatto and E.~Remiddi,
Nucl. Phys. B {\bf 95}, 93 (1980).

\bibitem{bodwin} G.T.~Bodwin, E.~Braaten, T.C.~Yuan and G.P. LePage,
Phys. Rev. D {\bf 46}, 3703 (1992).

\bibitem{matsuki} D.-Y.~Chen, J.~He, X.~Liu and T.~Matsuki,
arXiv:1207.3561 [hep-ph].

\bibitem{belle_kd0d0} J. Brodzicka {\etal} (Belle Collaboration),
Phys. Rev. Lett. {\bf 100}, 092001 (2008).

\bibitem{uehara} Sadaharu Uehara, private communication.

\bibitem{belle2} T.~Abe {\etal} (BelleII Collaboration), 
arXiv:1011.6352 [hep-ex]. 

\bibitem{belle_x3940} K.~Abe {\etal} (Belle Collaboration),
Phys. Rev. Lett. {\bf 98}, 082001 (2007).

\bibitem{belle_x4160} P. Pakhlov {\etal} (Belle Collaboration),
Phys. Rev. Lett. {\bf 100}, 202001 (2008).

\bibitem{chao} K.-T.~Chao, Phys. Lett. {\bf B661}, 348 (2008).

\bibitem{mizuk} Roman Mizuk, private communication.

\bibitem{hybrids} D.~Horn and J.~Mandula,
Phys. Rev. D {\bf 17}, 898 (1978);
F.~Close, I.~Dunietz, P.R.~Page, S.~Veseli and H.~Yamamoto,
Phys. Rev. D {\bf 57}, 5653 (1998);
G.~Chiladze, A.F.~Falk and A.A.~Petrov,
Phys. Rev. D {\bf 58}, 034013 (1998);
F.~Close and S.~Godfrey, Phys. Lett. {\bf B574}, 210 (2003).

\bibitem{diquarks} R.L.~Jaffe, Phys. Rev. D {\bf 15}, 267 (1977);
D.B.~Lichtenberg, W.~Namgung, E.~Predazzi and J.C.~Wills,
Phys. Rev. Lett. {\bf 48}, 1653 (1982);
L.~Maiani, F.~Piccinini, A.D.~Polosa and V.~Riquer,
Phys. Rev. D {\bf 71}, 014028 (2005);
L.~Maiani, F.~Piccinini, A.D.~Polosa and V.~Riquer,
Phys. Rev. D {\bf 72}, 031502(R) (2005);
L.~Maiani, A.D.~Polosa and V.~Riquer,
Phys. Rev. Lett. {\bf 99}, 182003 (2007)

\bibitem{molecules}
M.B.~Voloshin and L.B.~Okun,
JETP Lett. \textbf{23}, 333 (1976);
M. Bander, G.L.~Shaw and P.~Thomas,
Phys. Rev. Lett. \textbf{36}, 695 (1976);
A.~De~Rujula, H.~Georgi and S.L.~Glashow,  
Phys. Rev. Lett. \textbf{38}, 317 (1977);  
A.V. Manohar and M.B. Wise, Nucl. Phys. B \textbf{339}, 17 (1993);
N.A.~T{\" o}rnqvist, Z. Phys. C {\bf 61}, 526 (1994).

\bibitem{hadrocharmonium} S.~Dubynskiy and M.B.~Voloshin,
Phys. Lett. {\bf B666}, 344 (2008).

\bibitem{bugg} D.V.~Bugg, EPL {\bf 96}, 11002 (2011).

\bibitem{dudek} L.~Liu {\etal} (Hadron Spectrum Collaboration)
JHEP {\bf 07}, 126 (2012). 

\bibitem{maiani_z3900} L.~Maiani, V.~Riquer, R.~Faccini,
F.~Piccinini, A.~Pilloni, and A.D.~Polosa,
Phys. Rev. D {\bf 87}, 111102(R) (2013).

\bibitem{voloshin_z3900} M.B.~Voloshin,
Phys. Rev. D {\bf 87}, 091501 (2013).

\bibitem{molina_x3915} R.~Molina and E.~Oset,
Phys. Rev. D {\bf 80}, 114013 (2009).


\end{thebibliography}
\end{document}